\documentclass[10.5pt]{article}
\usepackage[hypertex]{hyperref}
\usepackage{graphicx} 
\usepackage{amsmath,amssymb,amsfonts,bm,cite} 

\def\a{\alpha} \def\b{\beta}   \def\d{\delta} \def\D{\Delta}    \def\th{\theta}    \def\L{\Lambda} \def\m{\mu} \def\n{\nu}     \def\r{\rho} \def\s{\sigma}  \def\t{\tau}       

\def\dg{\dagger}  \def\nn{\nonumber}

\newcommand{\meV}{ {\rm meV} }  \newcommand{\keV}{ {\rm keV} } \newcommand{\MeV}{ {\rm MeV} } \newcommand{\GeV}{ {\rm GeV} }

 \newcommand{\lsp}{ \left ( } \newcommand{\rsp}{ \right ) } \newcommand{\Lg}{\mathcal{L}}  \newcommand{\To}{\Rightarrow}   

\newcommand{\vev}[1]{ \langle {#1} \rangle }
  \newcommand{\Det}{{\rm Det}}

\newcommand{\Diag}[3]{ \begin{pmatrix} #1 & 0 & 0 \\ 0 & #2 & 0 \\ 0 & 0 & #3 \\\end{pmatrix}}

\begin{document}

\begin{flushright}
STUPP-20-243
\end{flushright}

\vskip 1.35cm

\begin{center}
{\Large \bf Almost exact diagonal reflection symmetries \\ and three-zero texture}

\vskip 1.2cm

Masaki J. S. Yang

\vskip 0.4cm

{\it Department of Physics, Saitama University, \\
Shimo-okubo, Sakura-ku, Saitama, 338-8570, Japan\\
}




\begin{abstract} 

In this paper, we consider a three-zero texture with diagonal reflection symmetries in the SM. 
The three-zero texture has two less assumptions  ($(M_{u})_{11} , (M_{\nu})_{11} \neq 0$) than the  universal four-zero texture of mass matrices $(M_{f})_{11} = (M_{f})_{13,31} =  0$ for $f = u,d,\nu, e$.
The texture allows diagonal reflection symmetries to be almost exact and $d$\,-\,$e$ unification. 
They reproduce the CKM and MNS matrices with accuracies of $O(10^{-4})$ and $O(10^{-3})$.
Some perturbative diagonalizations yield
two relations with good accuracy for quark mass and mixing. 
Although this calculation is done in a special basis, it is a general result in a sense, because other textures and generalized $CP$ symmetries exist by some weak basis transformation.

By assuming a $d$\,-\,$e$ unified relation ($M_{d} \sim M_{e}$),  
we obtain the lightest neutrino mass $m_{1} \simeq 2.12\, - \, 5.64\, [\meV]$ and 
the effective mass of the double beta decay $|m_{ee}| \simeq 1.23 - 2.15 \, [\meV]$.

\end{abstract} 

\end{center}

\section{Introduction}

Recently $CP$ violation of neutrino oscillations has been strongly suggested \cite{Abe:2018wpn}.
It sheds further light on flavor structures and the baryon asymmetry of the universe.
To explain the leptonic $CP$ violation,  
various generalized $CP$ symmetries (GCP) have been studied 
\cite{Ecker:1981wv, Ecker:1983hz, Gronau:1985sp, Ecker:1987qp,Neufeld:1987wa,Ferreira:2009wh,Feruglio:2012cw,Holthausen:2012dk,Ding:2013bpa,Girardi:2013sza,Nishi:2013jqa,Ding:2013hpa,Feruglio:2013hia,Chen:2014wxa,Ding:2014ora,Ding:2014hva,Chen:2014tpa,Chen:2015siy,Li:2015jxa,Turner:2015uta, Rodejohann:2017lre, Penedo:2017vtf,Nath:2018fvw}, 
including the $\m-\t$ reflection symmetry and its analog \cite{Harrison:2002et,Grimus:2003yn, Grimus:2005jk, Farzan:2006vj, Joshipura:2007sf, Adhikary:2009kz, Joshipura:2009tg, Xing:2010ez, Ge:2010js, He:2011kn, Gupta:2011ct, Grimus:2012hu, He:2012yt, Joshipura:2015dsa, Xing:2015fdg, He:2015afa, Chen:2015siy, He:2015xha, Samanta:2017kce, Xing:2017cwb, Nishi:2018vlz, Nath:2018hjx, Sinha:2018xof, Xing:2019edp, Pan:2019qcc, Yang:2020qsa}. 
Among them, GCPs called diagonal reflection symmetries (DRS) 
$R \, M_{u,\n}^{*} \, R = M_{u,\n}, ~ M_{d,e}^{*} = M_{d,e}$ with $R =$ diag $(-1,1,1)$ 
have been proposed \cite{Yang:2020goc}. 
By combining with the universal four-zero texture $(M_{f})_{11} = (M_{f})_{13,31} =  0$ for $f = u,d,\nu,e$ \cite{Fritzsch:1995nx, Xing:2015sva}, 
this system explains well all masses, mixings, and $CP$ violations of quarks and leptons. 
However, the previous study has two shortcomings. 
First, the DRS are broken with an accuracy of $2 \sim 3$\%.
Second, $d$\,-\,$e$ unification is difficult because the mass matrices of quarks and leptons have different hierarchies.

In order to improve these shortcomings, in this paper, we consider three-zero texture with DRS.
Compared to the universal four-zero texture,  
the three-zero texture has two fewer assumptions $((M_{u})_{11} \neq 0$ and $(M_{\nu})_{11} \neq 0)$.
With this texture, the diagonal reflection symmetries become almost exact ones, and 
 the CKM and MNS matrices are reproduced with high precision of $O(10^{-4})$ and $O(10^{-3})$.
In addition, this scheme allows a $d$\,-\,$e$ unified relation, which was difficult to achieve with the universal four-zero texture.

This paper is organized as follows. 
The next section gives a review of DRS.
In the third section, a realization of DRS and three-zero texture is discussed. 
In Sec.~4, we apply the DRS and three-zero texture to the lepton sector. 
The final section is devoted to a summary.

\section{Representation of CKM matrix, texture, and symmetries} 

First of all, we present a representation of the CKM matrix and 
the DRS in a previous study \cite{Yang:2020qsa, Yang:2020goc}. 
The mass matrices $M_{f}$ of the SM fermions $f= u,d,e,\n$ are given by
\begin{align}
\Lg \ni  \sum_{f=u,d,e} -  \bar f_{Li } M_{f ij} f_{Rj} - \bar \n_{L i} M_{\n ij} \n_{L j}^{c} + {\rm h.c.} \, .
\end{align}
The CKM and MNS matrices arise as a result of misalignment in the diagonalization of the two mass matrices, 
\begin{align}
 V_{\rm CKM} = U_{u}^{\dg} U_{d} , ~~~ 
 U_{\rm MNS} = U_{e}^{\dg} U_{\n} .  \label{mixings}
\end{align}
Here, $U_{f}$ is a unitary matrix that diagonalizes the mass matrix $M_{f}$.

There are several different ways to parameterize the CKM matrix.
Here we will consider the first parameterization by Fritzsch and Xing \cite{Fritzsch:1997fw}, 
\begin{align}
 V_{\rm CKM}   =   
\begin{pmatrix}
c_{u} & s_{u} & 0 \\
 - s_{u} & c_{u} & 0 \\
 0 & 0 & 1 \\
\end{pmatrix} 
\begin{pmatrix}
e^{- i \phi}  & 0 & 0 \\
 0 & c_{q} & s_{q} \\
 0 & - s_{q} & c_{q} \\
\end{pmatrix}
\begin{pmatrix}
c_{d} & - s_{d} & 0 \\
 s_{d} & c_{d} & 0 \\
 0 & 0 & 1 \\
\end{pmatrix} , 
\label{VCKM1}
\end{align}
where $s_{f} \equiv \sin \th_{f}, c_{f} \equiv \cos \th_{f}$. 
The physical advantage of this representation is that only $c_{q}$ and $s_{q}$ receive  renormalization.
Using the best fit of the recent data of the PDG parameterization \cite{Zyla:2020zbs}, 
\begin{align}
\sin \th_{12}^{\rm CKM} &= 0.22650, ~~ 
\sin \th_{23}^{\rm CKM} = 0.04053, ~~ 
\sin \th_{13}^{\rm CKM} = 0.00361, ~~
\d^{\rm CKM} = 1.196, 
\end{align}
we obtain magnitudes of these parameters as
\begin{align}
s_{u} & =  0.0887, ~~ s_{d} =  0.2100, ~~ 
 s_{q} = 0.04069, ~~ \phi = 88.66^{\circ} . 
\label{FZparameters}
\end{align}
In particular, even if we set $\phi = \pi/2$, these parameters reproduce the CKM matrix with an accuracy of $O(10^{-5})$. 
Thus, the unitary matrices $U_{u,d}$ can be interpreted as follows:
\begin{align}
U_{u} & =  
\begin{pmatrix}
+i  & 0 & 0 \\
 0 & c_{t} & s_{t} \\
 0 & - s_{t} & c_{t} \\
\end{pmatrix}
\begin{pmatrix}
c_{u} & -s_{u} & 0 \\
 s_{u} & c_{u} & 0 \\
 0 & 0 & 1 \\
\end{pmatrix} \label{Uu}
 , \\
U_{d} &= 
\begin{pmatrix}
1 & 0 & 0 \\
 0 & c_{b} & s_{b} \\
 0 & - s_{b} & c_{b} \\
\end{pmatrix}
\begin{pmatrix}
c_{d} & - s_{d} & 0 \\
 s_{d} & c_{d} & 0 \\
 0 & 0 & 1 \\
\end{pmatrix} ,  \label{Ud}
\end{align}
where $\sin (\th_{b} - \th_{t}) = \sin \th_{q}$. 
These parameters have degrees of freedom by redefinition the fields, 
{\it e.g.,} $\th_{t}' = \th_{t} + \th , ~ \th_{b}' = \th_{b} + \th.$
From Eqs.~(\ref{Uu}) and (\ref{Ud}) the mass matrices of quarks can be naturally reconstructed as 
\begin{align}
M_{u} &= U_{u} \, M_{u}^{\rm diag} \, U_{u}^{\dg}  \simeq 
\begin{pmatrix}
m_{u1} + s_{u}^2 m_{u2} & - i s_{u} m_{u2} & i s_{u} s_{t} m_{u2} \\
 i s_{u} m_{u2} & m_{u2} + s_{t}^{2} m_{u3} & s_{t} m_{u3} \\
- i s_{u} s_{t} m_{u2} & s_{t} m_{u3} &  m_{u3} 
\end{pmatrix} , \label{texture1} \\
M_{d} &= U_{d} \, M_{d}^{\rm diag} U_{d}^{\dg}  \simeq
\begin{pmatrix}
 m_{d1} + s_{d}^2 m_{d2} & - s_{d} m_{d2} & s_{d} s_{b} m_{d2}  \\
- s_{d} m_{d2} & m_{d2} + s_{b}^{2} m_{d3} & s_{b} m_{d3} \\
s_{d} s_{b} m_{d2}  & s_{b}  m_{d3} & m_{d3} 
\end{pmatrix} , 
\label{texture2}
\end{align}
where $M_{f}^{\rm diag} = {\rm diag} (m_{f1}, m_{f2}, m_{f3})$ is a diagonal matrix that has complex phases. 
The phases of mass eigenvalues are unphysical in the SM.
However, in this case, they affect physical quantities (for example, the Jarlskog invariant Eq.~(\ref{Jarls}))
and the shape of texture through the phase of mixing. 
Thus, we will consider them explicitly. 

The running masses at the weak scale $\mu = m_{Z}$ is given by \cite{Xing:2011aa}
\begin{align}
\begin{array}{lll}
m_{u} = 1.38^{+0.42}_{-0.41} \, [\MeV],  &m_{c} = 638^{+43}_{-84} \, [\MeV], &  m_{t} = 172.1  \pm 1.2 \, [\GeV],  \\[6pt]
m_{d} =  2.82 \pm 0.48 \, [\MeV], & m_{s} = 57^{+18}_{-12} \, [\MeV], & m_{b} = 2860^{+160}_{-60} \, [\MeV] .
\end{array} 
\label{masses}
\end{align}
These values indicate some facts about the textures~(\ref{texture1}) and (\ref{texture2}):
\begin{itemize}
\item A condition $m_{d1} \, m_{d2} <0$ realizes a zero texture $(M_{d})_{11} \simeq 0$ because of
$s_{d} \simeq \sqrt{m_{d} / m_{s}}$.
This is known as the GST relation \cite{Gatto:1968ss} and has been studied in various papers such as the four-zero texture  \cite{Fritzsch:1995nx}. 

\item Although $m_{u1} m_{u2} < 0$ holds, 
$s_{u} \simeq 0.09$ in Eq.~(\ref{FZparameters}) is too large to make a zero texture 
$(M_{u})_{11} = 0$. 
Conversely, a choice of the parameter $s_{u} \simeq \sqrt{m_{u} / m_{c}} \simeq 0.046$ 
predicts too small 13 element $|V_{ub}| \simeq |V_{cb}| s_{u} \simeq 0.0018$ \cite{ Roberts:2001zy}. 

\item To create the universal texture zero (UTZ) $(M_{f})_{11} = 0$ \cite{Albright:1989if,Rosner:1992qa,Roberts:2001zy, Koide:2002cj, Grimus:2004hf,deMedeirosVarzielas:2017sdv,deMedeirosVarzielas:2018vab}  by a basis transformation from Eq.~(\ref{texture1}), a 12 and/or 13 rotation for $M_u$ is required. 
However, these mixings break the DRS~(\ref{diagref}) because of a discussion under Eq.~(\ref{mutausym}). 

\item Each of the 13 elements is very small; $|(M_{u})_{13}| \simeq s_{u} s_{t} m_{c} \sim 0.57 \, [\MeV]$ and $|(M_{d})_{13}| \simeq s_{d} s_{b} m_{s} \sim  0.60 \, [\MeV]$ with 
$s_{b} \simeq 0.05$ and $s_{t} \simeq 0.01$. 
Even if these matrix elements $\lesssim$ 1\,MeV are replaced with $0$ by hand, 
they only produce errors of $O(10^{-4})$ in the masses and mixings.
\end{itemize}
From these facts, the following Hermitian matrices can
 reproduce the quark masses and mixings with high precision;
\begin{align}
M_{u}' = 
\begin{pmatrix}
D_{u} & i \, C_{u} & 0 \\ 
-i \, C_{u} &\tilde B_{u} & B_{u} \\
0 & B_{u} & A_{u} 
\end{pmatrix} , ~~~
M_{d}' =  
\begin{pmatrix}
0 & C_{d} & 0 \\ 
C_{d} &\tilde B_{d} & B_{d} \\
0 & B_{d}  & A_{d} 
\end{pmatrix} ,
\label{threezero}
\end{align}
where $A_{f} \sim D_{f}$ are real parameters which satisfy
\begin{align}
|A_{f}| \gg  |\tilde B_{f}| , |B_{f}| \gg |C_{f}| > |D_{f}| \, . 
\label{hier}
\end{align}
We call Eq.~(\ref{threezero}) as {\it three-zero texture}, 
following the same zero counting method by Fritzsch \cite{Fritzsch:1977vd}.
Although mass matrices are generally not Hermitian, 
$M_{f}$ are assumed to be Hermitian 
that are justified by the parity symmetry  in the left-right symmetric models \cite{Pati:1974yy,Senjanovic:1975rk,Mohapatra:1974hk}.  

\subsection{Three-zero texture and diagonal reflection symmetries}

Here, we will mention some properties of the three-zero texture and DRS.
The determinants of Eq.~(\ref{threezero}) are
\begin{align}
\Det [M_{u}'] & =  - A_{u} C_{u}^{2} +  A_{u} \tilde B_{u} D_{u}  - B_{u}^2 D_{u} , \\
\Det [M_{d}'] & = - A_{d} C_{d}^{2} \, . 
\end{align}
Using a freedom of overall sign, we can set $A_{u,d}, \, m_{t,b }> 0$ without loss of generality.
Since it yields a condition $\Det [ M_{d}'] <0$, one of the down-type mass eigenvalues $m_{d1}$ or $m_{d2}$ is always negative \cite{Xing:2015sva}. 
For up-type fermions, there are four possibilities $(m_{u1}, m_{u2}) = (\pm m_{u}, \pm m_{c})$  because the sign of determinant is indefinite. 

Analytical solutions for the diagonalization of $M_{d}$ and four-zero textures can be found in \cite{Xing:2015sva}. 
For $M_{u}$, an analytic form of the exact diagonalization is also found in Eq. (4.4) - (4.6) in a review of zero textures \cite{Fritzsch:1999ee}.
By removing the complex phase, the two mass matrices~(\ref{threezero}) can be diagonalized by real orthogonal matrices $O_{f}$. 
Moreover, 13 mixings of $O_{f}$ can be neglected due to the hierarchy~(\ref{hier}). Thus, the matrices $O_{f}$ can be approximately written by a product of two two-dimensional rotations \cite{Xing:2015sva}.
\begin{align}
O_f & \simeq 
\begin{pmatrix}
1 & 0 & 0 \\
0 & 1 & t_{f}  \\
0 & - t_{f}  & 1 
\end{pmatrix} 
\begin{pmatrix}
c_{f} &  s_{f} & 0 \\
-  s_{f} & c_{f} & 0 \\
  0 & 0 & 1
\end{pmatrix}
\simeq
\begin{pmatrix}
1 & s_{f} & 0 \\
-  s_{f} & 1 & t_{f}  \\
 s_{f} \, t_{f}& - t_{f} & 1 
\end{pmatrix} \, , 
\label{Bf}
\end{align}
where 
\begin{align}
t_{f} \simeq {B_{f} \over A_{f}}, ~~~ 
s_{u} \simeq {C_{u} \over \tilde B_{u}} , ~~~
s_{d} \simeq {C_{d} \over \tilde B_{d}} \simeq \pm \sqrt{m_{d} \over m_{s} + m_{d} } .
\end{align}
Note that the signs of $t_{f}, s_{f}$ are partially physical. As in Ref.~\cite{Xing:2015sva}, 
a basis with $C_{d} > 0$ relates the sign of mass eigenvalues and mixing angle because $ {\rm sign} ( C_{d} /\tilde B_{d}) =  {\rm sign} (\tilde B_{d}) =  {\rm sign} ( m_{d2} ) = - {\rm sign} ( m_{d1} )$. 

The CKM matrix $V_{\rm CKM} = U_{u}^{\dg} U_{d}$ is calculated as 
\begin{align}
V_{\rm CKM} &\simeq O_{u}^{T} \Diag{-i}{1}{1} O_{d} \simeq
\begin{pmatrix}
- i & - s_{u} & s_{u} \, t_{u} \\
- i  s_{u} & 1 & - t_{u}  \\
0 &  t_{u} & 1 
\end{pmatrix} 
\begin{pmatrix}
1 & s_{d} & 0 \\
- s_{d} & 1 & t_{d}  \\
 s_{d} \, t_{d}& - t_{d} & 1 
\end{pmatrix} 
\label{VCKM2}
\\ & \simeq 
\begin{pmatrix}
-i  & - i s_{d} - s_{u}  & s_{u} (t_{u} - t_{d}) \\
- i s_{u} - s_{d} & 1 & t_{d} - t_{u} \\
s_{d}  (t_{d} - t_{u})  &  t_{u} - t_{d} & 1 \\
\end{pmatrix} .
\end{align}
The mass matrices~(\ref{threezero}) has nine parameters. At leading order of the perturbation, one of $t_{u}$ or $t_{d}$ (equivalently, $B_{u}$ or $B_{d}$) is not determined.
Instead, the six quark masses and other two input parameters $V_{cb} = t_{d} -t_{u}$ and ${V_{ub} / V_{cb}} = s_{u}$ yield two non-trivial relations for $s_{d}$ and the Jarlskog invariant \cite{Jarlskog:1985ht}, 
\begin{align}
| s_{d} | &\simeq \left|{V_{td} \over V_{ts}} \right| = 0.215 \, ,  \simeq \sqrt{m_{d} \over m_{d} + m_{s} }  
\simeq 0.217 \, , 
\\
J_{\rm CKM} &= {\rm Im} \, [ V_{1 1} V_{3 3} V_{1 3}^{*} V_{3 1}^{*}  ] 
= 3.00^{+0.15}_{-0.09} \times 10^{-5} , 
\\ & 
=  c_{u} s_{u} c_{d} s_{d} c_{q} s_{q}^{2} \simeq  \sqrt{m_{d} \over m_{d} + m_{s} }
\, { \left| V_{ub} \over V_{cb} \right| } | V_{ts}|^{2}
\simeq  3.06 \times 10^{-5} ,
\label{Jarls}
\end{align}
where $s_{q} \simeq t_{u} - t_{d}$.
These two relations hold with an accuracy of $1\sim2\,\%$.
In Eq.~(\ref{VCKM1}), $c_{u}, c_{d}, c_{q}$ can all be taken to be positive by redefinitions of the quark fields \cite{Fritzsch:1997fw}. In such a basis, ${\rm sign} (s_{u} s_{d}) >0$ is required to keep  $J_{\rm CKM}$ positive. 

Meanwhile, the mass matrices~(\ref{texture1}) and (\ref{texture2}) 
satisfy the diagonal reflection symmetries \cite{Yang:2020goc};
\begin{align}
R \, M_{u}^{*} \, R = M_{u}, ~~ M_{d}^{*} = M_{d}.  ~~
 R \equiv \Diag{-1}{1}{1}. 
 \label{diagref}
\end{align}
These symmetries concentrate maximal complex phases only in the first generation.
It suggests that flavored $CP$ phases accompany the chiral symmetry breaking of the first generation and the Higgs boson gives ``special treatment'' to the first generation.

The three-zero texture makes the DRS almost exact symmetries for the best-fit values of the quarks. 
Moreover, these symmetries are hardly renormalized in SM 
because the coupling constants of the first generation are very small \cite{Yang:2020goc}. 
As a result, in any renormalization scale, these zero textures and symmetries are a quite good description of the quark mass matrices.

Since this system has only nine parameters, 
Hermitian quark mass matrices can be reconstructed from 
six mass eigenvalues and the CKM matrix. 
As an example, 
the mass eigenvalues (\ref{masses}) and the following four input values 
\begin{align}
s_{d} > 0 , ~~ 
\left| {V_{ub} \over V_{cb}} \right| = s_{u} \simeq {C_{u} \over \tilde B_{u} - D_{u}} > 0, ~~ 
{B_{d} \over A_{d}} = {5 \over 4} |V_{cb}|,  ~~ 
{B_{u} \over A_{u}} = {1 \over 4} |V_{cb}|, 
\label{inputs}
\end{align}
reproduce the mass matrices as 
\begin{align}
M_{u}' = 
\begin{pmatrix}
 -3.6122 &  54.810 \, i & 0 \\
 - 54.810 \, i & -596.29 & 1752.8 \\
 0 & 1752.8 & 171682 \\
\end{pmatrix} \, [\MeV],  ~~~ 
M_{d}' = 
\begin{pmatrix}
 0 & 12.282 & 0 \\
 12.282 & - 46.742 & 149.56 \\
 0 & 149.56 & 2882.4 \\
\end{pmatrix} \, [\MeV] .  
\label{exmass}
\end{align}
Since $B_{u,d}$ cannot be determined at leading order of the perturbation (\ref{VCKM2}),  
their values in Eq.~(\ref{inputs}) that satisfy ${B_{d} \over A_{d}} - {B_{u} \over A_{u}} = |V_{cb}|$ 
are chosen by hand. 
By these choices, relations $B_{u} \simeq 3 m_{c} \simeq - 3 \tilde B_{u}$ and $B_{d} \simeq 3 m_{s} \simeq - 3 \tilde B_{d}$ hold in $M_{u,d}'$~(\ref{exmass}). 
Such textures can be realized in a $SU(3)_{F}$ model by 
a coupling ${1\over \L^{2}}\bar q \, \phi_{23}\, \phi_{23}^{\dg}\, q' H $ 
to a three-representation flavon $\phi_{23}$ that acquires a vacuum expectation values (vevs) 
 $\vev{\phi_{23}}\propto (0, 1, -3)$. 

A reconstructed CKM matrix is found to be 
\begin{align}
|V_{\rm CKM}'| \equiv 
| U_{u}'{}^{\dg} U_{d}' | = 
\begin{pmatrix}
0.97413 & 0.22597 & 0.00362 \\
0.22584 & 0.97332 & 0.04054 \\
0.00832 & 0.03984 & 0.99917
\end{pmatrix} ,
\end{align}
where $U_{f}'$ diagonalizes the mass matrices~(\ref{exmass}). 
The mixing matrix obtained by this procedure has an error of only $O(10^{-4})$ and has precise accuracy;
\begin{align}
|V_{\rm CKM}| - |V_{\rm CKM}'| = 
\begin{pmatrix}
1.2 & -5.3 & 0.1 \\
-5.3 & 1.2 & 0.1 \\
-2.3 & 0.6 & 0.0
\end{pmatrix}
\times 10^{-4} .
\label{error1}
\end{align}
If we replace the element $(M_{u}')_{11}$ in Eq.~(\ref{exmass}) with zero by hand, these errors will be about $10^{-3}$, and the accuracy of $V_{ub}, V_{td} \sim O(10^{-3})$ will be quite poor.

\subsection{Weak basis transformation and GCP}

At first glance, these results appear to be based on a special basis.
However, at least in the quark sector, there are other (almost exact) GCPs even if the basis is changed by a weak basis transformation (WBT) \cite{Branco:1999nb,Branco:2007nn}.

A WBT is a transformation of the following form, 
\begin{align}
& M_{u} \to M_{u}' = W^{\dg}_{q} M_{u} W_{u} \, , \nn \\  
& M_{d} \to M_{d}' = W^{\dg}_{q} M_{d} W_{d} \, ,  
\label{WBT}
\end{align}
that does not change the CKM matrix; 
\begin{align}
V^{\prime \prime}_{\rm CKM} \equiv U_{u}^{\dg} W_{q}^{\dg} W_{q} U_{d} = U_{u}^{\dg} U_{d} = V_{\rm CKM} \, . 
\end{align}
In particular, Hermiticity of $M_{u,d}'$ restricts $W_{f}$ to be $W_{q} =W_{u} = W_{d}$. 
In any basis transformed by a WBT, the DRS~(\ref{diagref}) can be deformed into other GCPs as follows
\begin{align}
L_{u}^{*} M_{u}^{\prime *} R_{u} = M_{u}' ,  ~~~
L_{d}^{*} M_{d}^{\prime *} R_{d} = M_{d}'  .
\label{GCP}
\end{align}
Here, 
\begin{align}
&L_{u} = W_{q}^{T} R \, W_{q} \, , ~~ L_{d} = W_{q}^{T}  W_{q} \, , \\
&R_{u} = W^{T}_{u} R \, W_{u}  \, , ~~ R_{d} = W^{T}_{d} W_{d} \, . 
\label{GCP2}
\end{align}
Since they are just equivalence transformations, these GCPs~(\ref{GCP}) (with deformed three-zero texture) also reproduce the CKM matrix with an accuracy of about $O(10^{-4})$ and are almost renormalization invariant.
As a concrete example, the following unitary matrix
\begin{align}
W_{q} = W_{u} = W_{d} = 
U_{BM} \equiv
\begin{pmatrix}
 1 & 0 & 0 \\
 0 & \frac{i}{\sqrt{2}} & \frac{i}{\sqrt{2}} \\
 0 & -\frac{1}{\sqrt{2}} & \frac{1}{\sqrt{2}} \\
\end{pmatrix} ,  ~~~
\end{align}
generates new GCPs with
\begin{align}
U_{BM}^{T}  R \, U_{BM} \equiv - T_{u} = %
\begin{pmatrix}
 -1 & 0 & 0 \\
 0 & 0 & -1 \\
 0 & -1 & 0 \\
\end{pmatrix} , ~~~
U_{BM}^{T} U_{BM} \equiv T_{d} = 
\begin{pmatrix}
 1 & 0 & 0 \\
 0 & 0 & -1 \\
 0 & -1 & 0 \\
\end{pmatrix} .
\end{align}
These operators produce two separate $\m-\t$ reflection symmetries \cite{Yang:2020qsa}; 
\begin{align}
T_{u} M_{u}^{\prime *} T_{u} = M_{u}'  , ~~~ 
T_{d} M_{d}^{\prime *} T_{d} = M_{d}' .
\label{mutausym}
\end{align} 

WBTs that preserve DRS satisfies $L_{u} = R_{u} = R, ~ L_{d} = R_{d} = 1$ and thus have the following properties.
First, $W_{d}$ is an orthogonal matrix $O_{d}$.
In the same way, $W_{u}$ is found to be $W_{u} = P^{*} O_{u} P$ with $P = {\rm diag} ( i ,1,1)$.
Since $W_{q}$ needs to satisfy these two conditions simultaneously, only a real 23 rotation  is allowed.

As a result, under the freedom of WBT, the three-zero texture and DRS~(\ref{threezero}) cover the general situation of quark mass matrices.
In the lepton sector, the mixing matrix $U_{\rm MNS}'$ we see in section 4 is also a general result under WBT. 
However, since the lightest neutrino mass $m_{1}$ and the Majorana phases $\a_{2,3}$ has not been determined, different physical observables lead to mass matrices that are physically inequivalent from Eqs.~(\ref{mn1}) and (\ref{mn2}).

\section{Realization of the textures and symmetries }

The realization of zero textures by discrete symmetry or continuous symmetry 
has been discussed in several studies \cite{Weinberg:1977hb,Fritzsch:1977vd, Grimus:2004hf}. 
Meanwhile, $\m-\t$ reflection symmetry naturally appear from GCP and $S_{4}$ symmetry \cite{Mohapatra:2012tb, Feruglio:2012cw, Ding:2013hpa}. 
Because DRS are equivalent to two different $\m-\t$ symmetries, 
a similar model building would be possible.

The DRS are inconsistent with the gauge symmetry of the SM because the left-handed fermions transform separately. However, they can be a remnant subgroup of a larger CP symmetry.
Such residual GCPs have been discussed in Ref.~\cite{Ding:2013bpa}. 
In order to realize the DRS and zero textures in a field theory,
 the original paper \cite{Yang:2020goc} assumed $U(1)_{\rm PQ}\times Z_{2}\times {\rm GCP}$ symmetries in the two Higgs doublet model (2HDM). 
The $U(1)_{\rm PQ}$ symmetry produces a {\it flaxion} \cite{Ema:2016ops} or {\it axiflavon} \cite{Calibbi:2016hwq}.  
In this argument, a small Yukawa coupling $(Y_{u})_{11} = O(10^{-5}) \neq 0$ can arise from higher order effects of flavons.
In this section, we will see how the three-zero texture and DRS are actually constructed.

In this model, two SM singlet flavon fields $\th_{u, d}$ are introduced to the 2HDM. 
Fields and charge assignments in the model are represented in Table 1.
These flavons have nontrivial charges under the $U(1)_{\rm PQ}$ and GCP symmetries. 
Simultaneous breaking of these symmetries by vevs of $\th_{u,d}$ induces GCPs that are not well-defined in the SM. 
\begin{table}[h]
  \begin{center}
    \begin{tabular}{|c|ccccc|} \hline
           & $SU(2)_{L}$ & $U(1)_{Y}$ &  $Z_{2}^{\rm NFC}$ & $U(1)_{\rm PQ}$ & GCP \\ \hline \hline
      $q_{Li}$ & \bf 2 & $1/6$ & 1 & $-1,0,0$ & 1 \\
      $u_{Ri}$ & \bf 1 & $2/3$ & 1  & $1,0,0$ & 1\\ 
      $d_{Ri}$ & \bf 1 & $-1/3$ & $-1$ & $1,0,0$ & 1\\ 
      $l_{Li}$ & \bf 2 & $-1/2$ & 1 & $-1,0,0$ & 1\\
      $\n_{Ri}$ & \bf 1 & $0$ & 1 & $1,0,0$ & 1\\ 
      $e_{Ri}$ & \bf 1 & $-1$ &$-1$ & $1,0,0$ &  1 \\ \hline
      $H_{u}$ & \bf 2 & $-1/2$ & 1 & 0 & 1\\ 
      $H_{d}$ & \bf 2 & $1/2$ & $-1$ & 0  & 1 \\       
      $\th_{u}$ & \bf 1 & $1$ & 1 & $-1$ & $+i$ \\  
      $\th_{d}$ & \bf 1 & $1$ & $-1$ & $-1$ & $-i$\\ \hline
    \end{tabular}
    \caption{Charge assignments of fields under gauge, flavor, and GCP symmetries,}
  \end{center}
\end{table}
%

Under the $U(1)_{\rm PQ}$ symmetry, 
only the first-generation has nontrivial charges as
\begin{align}
f_{1 L}  \to e^{-i \a} f_{1L}, ~~ 
f_{1 R} &\to e^{i \a} f_{1R} , 
\end{align}
where $f_{L} = q_{L}, l_{L}$ and $f_{R} = u_{R}, d_{R}, \n_{R}$, and $e_{R}$. 
The bilinear terms $\bar f_{Li} f_{Rj}$ associated with Yukawa interactions are transformed as
\begin{align}
\lsp
\begin{array}{c|cc}
e^{ 2 i \a} & e^{i \a} & e^{i \a} \\ \hline
e^{i \a} & 1 & 1 \\
 e^{i \a} & 1 & 1 \\
\end{array}
\rsp .
\end{align}

Under these symmetries, 
the most general Yukawa interactions for quarks are written as
\begin{align}
- \Lg &\ni  \bar q_{L} (\tilde Y_{u}^{0} + {\th_{u} \over \L} \tilde Y_{u}^{1} + {\th_{u}^{2} \over \L^{2}} \tilde Y_{u}^{2} + {\th_{d}^{2} \over \L^{2}} \tilde Y_{u}'{}^{2}  ) u_{R}  H_{u} \label{37}
\\ & +  \bar q_{L} (\tilde Y_{d}^{0} + {\th_{d} \over \L} \tilde Y_{d}^{1} + {\th_{u} \th_{d} \over \L^{2}} \tilde Y_{d}^{2} )d_{R} H_{d} + h.c. \, ,  \label{38}
\end{align}
where $\L$ is a cut-off scale. Similar formulae hold for leptons.
The Yukawa matrices are parameterized as
\begin{align}
\tilde Y_{u,d}^{0} =
\begin{pmatrix}
0 & 0 & 0 \\ 
0 & \tilde d_{u,d} & \tilde c_{u,d} \\
0 & \tilde b_{u,d} & \tilde a_{u,d}
\end{pmatrix} ,  ~~~ 
\tilde Y_{u,d}^{1}  =
\begin{pmatrix}
0 & \tilde e_{u,d} & \tilde f_{u,d} \\ 
\tilde g_{u,d} & 0 & 0 \\
\tilde h_{u,d} & 0 & 0
\end{pmatrix} ,  ~~~ 
\end{align}
and $\tilde Y^{2}_{f}$ have only an 11 matrix element. 
From the charge assignments of GCP in Table 1, 
the generalized $CP$ invariance is defined as 
\begin{align}
\th_{u}^{*} = +i \th_{u}, ~~ \th_{d}^{*} = - i \th_{d} , ~~ \phi^{*} = \phi ~~ 
\text{for other bosonic and fermionic fields.}
\label{CPcharge}
\end{align}
This kind of old-fashioned GCP has been discussed in original papers \cite{Ecker:1983hz, Ecker:1987qp}.
It restricts relative complex phases of the matrix elements as
\begin{align}
& \tilde Y_{u,d}^{0} = \pm |\tilde Y_{u,d}^{0}| ,  ~~~
\tilde Y_{u}^{1} = \pm e^{i \pi/4} |\tilde Y_{u}^{1}| ,  ~~~ 
\tilde Y_{d}^{1} = \pm e^{- i \pi/4} |\tilde Y_{d}^{1}| , 
\label{yukawaphases}  \\
& \tilde Y_{u}^{2} = \pm i |\tilde Y_{u}^{2}| , ~~~ 
\tilde Y_{u}'{}^{2} = \pm i |\tilde Y_{u}'{}^{2}| , ~~~ 
\tilde Y_{d}^{2} = \pm |\tilde Y_{d}^{2}| .
\label{yukawaphases2}
\end{align}

It is easy to see that this GCP satisfies the consistency condition \cite{Feruglio:2012cw, Holthausen:2012dk}. 
The condition is defined as follows
\begin{align}
X \r (g)^{*} X^{-1} = \r (g') , ~~ g' \in G_{F}, 
\end{align}
where $\rho(g)$ is a representation of a flavor symmetry $G_{F}$ 
 that represents a transformation of a field $\Phi$, 
$\Phi \to \rho (g) \Phi, g \in G_{F}$. 
A matrix $X$ represents a GCP transformation $\Phi \to X \Phi^{*}$.
Because all flavor and GCP symmetries in this model are Abelian, 
$X$ and $\r (g)$ are commutative, so it leads to the usual $CP$ relation $\rho (g') = \rho (g)^{*}$. 
Since  $(e^{i \a})^{*} = e^{- i \a}\in U(1)_{\rm PQ}$ holds, 
 this GCP is well-defined. 

The scalar potential can be written as 
\begin{align}
V = V^{1}(H_{u}, H_{d}) + V^{2}(H_{u,d}, \th_{u,d}) + V^{3}(\th_{u}, \th_{d}).
\end{align}
Under the symmetries of the model, the whole scalar potential $V$ has only real terms \cite{Yang:2020goc}. Thus, 
real vevs of the flavon fields $\vev{\th_{u,d}}$ 
cause spontaneous symmetry breakings (SSB) other than gauge symmetries. 
The $CP$ phases are concentrated only on the first generation of Yukawa matrices in this basis. 

As a result, 
the vevs $\vev{\th_{u,d}}$  produce the following Yukawa matrices 
\begin{align}
Y_{u,d} = (\tilde Y_{u,d}^{0} + {\vev{\th_{u,d}}\over \L} \tilde Y_{u,d}^{1}
 + {\vev{\th_{u,d}}^{2} \over \L^{2}} \tilde Y_{u,d}^{(\prime)}{}^{2}) 
= 
\begin{pmatrix}
O({\vev{\th_{u,d}}^{2} \over \L^{2}} )  & e_{u,d} \, {\vev{\th_{u,d}}\over \L} e^{i \varphi_{u,d}} & f_{u,d} {\vev{\th_{u,d}}\over \L} \, e^{i \varphi_{u,d}} \\[3pt]
g_{u,d} \, {\vev{\th_{u,d}}\over \L} e^{i \varphi_{u,d}} &  d_{u,d} & c_{u,d} \\[3pt]
h_{u,d} \,{\vev{\th_{u,d}}\over \L} e^{i \varphi_{u,d}} & b_{u,d} & a_{u,d}
\end{pmatrix} ,
\label{fullYukawa}
\end{align}
where
\begin{align}
\varphi_{u} = + \pi/4 , ~~~ \varphi_{d} = - \pi/4 .
\label{phases}
\end{align}
with real parameters $a_{u,d} \sim g_{u,d}$.

We show below that Eq.~(\ref{fullYukawa}) becomes a three-zero texture with DRS (\ref{threezero}) by proper WBTs. 
This kind of realization of zero textures by WBT has been used 
in early studies of general parity and $CP$ symmetry such as \cite{Ecker:1980rw, Ecker:1980at, Ecker:1981wv, Ecker:1983hz}. 
First, 11 elements $(Y_{u})_{11} \neq 0$ and $(Y_{d})_{11} \simeq 0$ are realized by higher order effects. 
The vevs of flavons can be estimated from the
reconstructed values for $M_{u,d}'$~(\ref{exmass}) as 
\begin{align}
 {\vev{\th_{u}} \over \L} | \tilde Y_{u}^{1} | &\sim {\sqrt{2} \, C_{u}  \over v \, \sin \b}  \simeq {3 \times 10^{-4} \over \sin \b} ,  \label{flavonvev1} \\
  {\vev{\th_{d}} \over \L} | \tilde Y_{d}^{1} | &\simeq {\sqrt{2 m_{d} \, m_{s}} \over v \, \cos \b}  \simeq {0.7 \times 10^{-4} \over \cos \b} ,
 \label{flavonvev2}
 \end{align}
where $\vev{ H_{u}^{0}} \equiv v \sin \b / \sqrt 2,  \vev{H_{d}^{0}} \equiv v \cos \b / \sqrt 2 $ with $\vev{H_{u}^{0}}^{2} + \vev{ H_{d}^{0}}^{2} = v^{2}/2$.  
The grand unified scale $\L_{\rm GUT} \sim 10^{16}$ GeV 
suggests magnitudes of vevs as $\vev{\th_{u,d}}\sim \L_{\rm GUT} \sqrt{m_{u,d} m_{c,s}} / v   \sim 10^{12}$ GeV.
From Eq.~(\ref{flavonvev2}), the vev $\vev{\th_{d}}$ can be large for large $\tan \b \gg 1$. 
In this case, the coupling $\tilde Y_{u}'{}^{2}$ generates a small 11 element for $Y_{u}$~(\ref{fullYukawa}). 
It is estimated as 
\begin{align}
(Y_{u})_{11} \simeq  {\vev{\th_{d}}^{2} \over \L^{2}} \tilde Y_{u}'{}^{2}
 \simeq \pm i \, {10^{-8} \, \tan^{2} \b \over | \tilde Y_{d}^{1} |^{2} } | \tilde Y_{u}'{}^{2} |
\sim \pm i \times 10^{-5} ~~ {\rm for} ~~ \tan \b \simeq 30 , ~ {\tilde Y_{u}'{}^{2} \over |Y_{d}^{1}|^{2}} \sim 1.
\end{align}
%
%
For the down-type quarks, since $(Y_{d})_{11} \simeq {\vev{\th_{u}} \vev{\th_{d}} \over \L^{2}} \tilde Y_{d}^{2} $ is suppressed by $\tan \b$ than $(Y_{u})_{11}$, the zero texture approximation $(Y_{d})_{11} \simeq 0$ is valid. 

Next, DRS and three-zero texture are realized by proper WBTs. 
Eq.~(\ref{fullYukawa}) with Eq.~(\ref{phases}) approximately satisfies deformed DRS~(\ref{GCP}) 
\begin{align}
R_{u} Y_{u}^{*} R_{u} = Y_{u} , ~~~ 
R_{d} Y_{d}^{*} R_{d} = Y_{d} ,
\label{refsym2}
\end{align}
where
\begin{align}
R_{u} = \Diag{+i}{1}{1} ,  ~~~ 
R_{d} = \Diag{-i}{1}{1} . 
\label{47}
\end{align}
A small symmetry breaking effect comes from 
the term $ {\th_{u} \th_{d} \over \L^{2}} \tilde Y_{d}^{2}$ in the 11 element of $Y_{d}$. 
Eq.~(\ref{47}) is equivalent to the original DRS~(\ref{diagref}) by the following WBT~(\ref{GCP}) with Eq.~(\ref{GCP2}).
\begin{align}
R_{u} = W^{T}_{u} R \, W_{u}  \, , ~~ R_{d} = W^{T}_{d} 1 W_{d} \, , ~~~
W_{u} = W_{d} =  \Diag{e^{- i \pi /4}}{1}{1} \, .  
\end{align}
This WBT is caused by a phase redefinition of the first generation as follows
\begin{align}
(u_{R1} , d_{R1}, \bar q_{L1} ) = e^{ i\pi/4} (u_{R1}' , d_{R1}' , \bar q_{L1}' ) \, .
\end{align}
This redefinition makes the GCP charge of the first generation equal to $\pm i$.
\begin{align}
(u_{R1}' , d_{R1}', \bar q_{L1}' )^{*} = i (u_{R1}' , d_{R1}' , \bar q_{L1}' ) \, , 
\label{1stGCP}
\end{align}
Further phase redefinition of $\th_{u}$ allows us to push the complex phases to only the vev of $\th_{u}$;
\begin{align}
\th_{u} = i \th_{u}' , ~~~ 
\vev{\th_{u}} = i \vev{\th_{u}'} . 
\end{align}
The GCP~(\ref{CPcharge}) for $\th_{u}$ is transformed as follows, 
\begin{align}
(\th_{u}')^{*} =  -  i \th_{u}' \, , ~~~ 
(\th_{d})^{*} = - i \th_{d} \, ,
\label{thGCP}
\end{align}
so that $u-d$ unification is possible.
In this basis, since the GCP charges of Eq.~(\ref{1stGCP}) and Eq.~(\ref{thGCP}) cancel, all the Yukawa matrices in Eqs.~(\ref{37}) and (\ref{38}) become real matrices and do not have complex phases as considered in Eqs.~(\ref{yukawaphases}) and (\ref{yukawaphases2}). 

Finally, 
we assume a $u-d$ unified relation $\tilde Y_{u}^{1} = \tilde Y_{d}^{1}$ in the basis of Eq.~(\ref{thGCP}). 
This $u-d$ unification also indicates the left-right symmetric model. 
A WBT (\ref{WBT}) by real 23 rotations $O_{23}$ and $O_{23}'$ realizes zero textures 
\begin{align}
(Y_{u})_{13} = (Y_{d})_{13} = (Y_{u})_{31} = (Y_{d})_{31} = 0.
\label{13zero}
\end{align}
Note that $O_{23}$ commutes with DRS $R \, O_{23}^{*} \, R = O_{23}$ because of the discussion under Eq.~(\ref{mutausym}).  
Explicitly, this step is written as
\begin{align}
O_{23}' Y_{u,d} O_{23} &= 
\begin{pmatrix}
1 & 0 & 0 \\
0 & c_{y'} & s_{y'} \\ 
0 & - s _{y'} & c_{y'}
\end{pmatrix}
\begin{pmatrix}
(Y_{u,d})_{11}  & e \, {\vev{\th_{u,d}}\over \L} e^{i \phi_{u,d}} & f {\vev{\th_{u,d}}\over \L} \, e^{i \phi_{u,d}} \\[3pt]
g \, {\vev{\th_{u,d}}\over \L} e^{i \phi_{u,d}} &  d_{u,d} & c_{u,d} \\[3pt]
h \,{\vev{\th_{u,d}}\over \L} e^{i \phi_{u,d}} & b_{u,d} & a_{u,d}
\end{pmatrix} 
\begin{pmatrix}
1 & 0 & 0 \\
0 & c_{y} & - s_{y} \\ 
0 & s _{y} & c_{y}
\end{pmatrix}
\\
& = 
\begin{pmatrix}
(Y_{u,d})_{11}  &  \sqrt{e^{2} + f^{2}} \, {\vev{\th_{u,d}}\over \L} \, e^{i \phi_{u,d}} & 0 \\[3pt]
\sqrt{g^{2} + h^{2}} \, {\vev{\th_{u,d}}\over \L}  \, e^{i \phi_{u,d}} & d'_{u,d} & c'_{u,d} \\[3pt]
0 & b'_{u,d} & a'_{u,d}
\end{pmatrix}  ,
\end{align}
where
\begin{align}
{s_{y} \over c_{y}} = {f \over e} , ~~ {s_{y'} \over c_{y'}} = {h \over g} , ~~ 
(Y_{d})_{11} \simeq 0 , ~~ \phi_{u} = \pi/2 , ~~~ \phi_{d} = 0 .
\end{align}
In this way, the DRS and three-zero texture is naturally realized by 
vev of GCP-charged flavons. 
Similar realizations have been discussed in the context of 
a sequential breaking of $U(2)$ flavor symmetry
\cite{Pomarol:1995xc,Barbieri:1995uv,Carone:1997qg,Barbieri:1997tu,Barbieri:1996ww,Blazek:1999ue,Barbieri:1998em,Dermisek:1999vy,Blazek:1999hz,Aranda:2001rd,Dudas:2013pja,Linster:2018avp,Linster:2020fww}. 

\section{Lepton sector}

In this section, we consider the following three-zero texture with DRS for the lepton sector. 
\begin{align}
M_{\n}' = 
\begin{pmatrix}
D_{\n} & i \, C_{\n} & 0 \\ 
i \, C_{\n} &\tilde B_{\n} & B_{\n} \\
0 & B_{\n} & A_{\n} 
\end{pmatrix} , ~~~
M_{e}' =  
\begin{pmatrix}
0 & C_{e} & 0 \\ 
C_{e} &\tilde B_{e} & B_{e} \\
0 & B_{e}  & A_{e} 
\end{pmatrix} .
\label{threezero2}
\end{align}
Here, $A_{f} \sim D_{f}$ are  real parameters. 
For $M_{e}'$, a hierarchy similar to that of quarks (\ref{hier}) is assumed as
\begin{align}
|A_{e}| \gg  |\tilde B_{e}| , |B_{e}| \gg |C_{e}| .
\label{hier2}
\end{align}
However, such a hierarchy is not imposed on $M_{\n}'$.
As with Eq.~(\ref{Bf}), a proper removal of the complex phase makes 
the diagonalization matrices of Eq.~(\ref{threezero2}) real orthogonal ones, 
\begin{align}
O_{\n} &\simeq 
\begin{pmatrix}
1 & 0 & 0 \\
 0 & c_{\n_{\t}} & s_{\n_{\t}} \\
 0 & - s_{\n_{\t}} & c_{\n_{\t}} \\
\end{pmatrix}
\begin{pmatrix}
c_{l'} & 0 & s_{l'} \\
 0 & 1 & 0 \\
- s_{l'} & 0 & c_{l'} \\
\end{pmatrix}
\begin{pmatrix}
c_{\n} & s_{\n} & 0 \\
-  s_{\n} & c_{\n} & 0 \\
 0 & 0 & 1 \\
\end{pmatrix} , \label{Un} \\
O_{e} & \simeq 
\begin{pmatrix}
 1 & 0 & 0 \\
 0 & c_{\t} & s_{\t} \\
 0 & - s_{\t} & c_{\t} \\
\end{pmatrix}
\begin{pmatrix}
c_{e} &s_{e} & 0 \\
- s_{e} & c_{e} & 0 \\
 0 & 0 & 1 \\
\end{pmatrix}  , 
\label{Ue}
\end{align}
where 
\begin{align}
s_{e} \simeq {C_{e} \over \tilde B_{e}} \simeq \pm \sqrt{m_{e} \over m_{\m} + m_{e}} \simeq \pm 0.07, 
~~~
s_{\t} \simeq {B_{e} \over A_{e}} . 
\end{align}
Note that the sign of $s_{\n}$ is opposite between Eq.~(\ref{VCKM2}) and (\ref{Un}), 
to match that of PDG parameterization. 
Due to the hierarchy~(\ref{hier2}), the 13 mixing of $O_{e}$ is neglected.
Since such a strong hierarchy is not assumed for neutrinos, 
$O_\n$ is taken to be the most general orthogonal matrix.
By neglecting the 13 mixing in $O_{e}$, an error that appears in $U_{e3}$ is less than 1\%; 
\begin{align}
{(M_{e}')_{13} \over (M_{e}')_{33}} \simeq
{\sqrt{m_{e} m_{\m}} \over m_{\t}} s_{\t} \simeq 0.004 \, s_{\t}, ~~
{0.004 \, s_{\t} \over \sin \th_{13}^{\rm MNS}} \simeq 0.00278~~ {\rm for} ~~s_{\t} = 0.1 \, .
\end{align}
This is negligible compared to the errors (\ref{35}) of the other inputs.

Similar to the CKM matrix~(\ref{VCKM2}), the MNS matrix is represented by two orthogonal matrices $O_{\n}$ and $O_{e}$; 
\begin{align}
 U_{\rm MNS}' &= 
U_{e}^{\dg} U_{\n} 
 = O_{e}^{T} \Diag{+i}{1}{1} O_{\n}
\\ & \simeq 
\begin{pmatrix}
c_{e} & - s_{e} & 0 \\
s_{e} & c_{e} & 0 \\
 0 & 0 & 1 \\
\end{pmatrix} 
\begin{pmatrix}
+ i & 0 & 0 \\
 0 & c_{l} & s_{l} \\
 0 & - s_{l} & c_{l} \\
\end{pmatrix}
\begin{pmatrix}
c_{l'} & 0 & s_{l'} \\
 0 & 1 & 0 \\
- s_{l'} & 0 & c_{l'} \\
\end{pmatrix}
\begin{pmatrix}
c_{\n} & s_{\n} & 0 \\
-  s_{\n} & c_{\n} & 0 \\
 0 & 0 & 1 \\
\end{pmatrix} , 
\label{UMNS1}
\end{align}
where $\sin (\th_{\n_{\t}} - \th_{\t}) = \sin \th_{l}$. 
The sign of phase corresponds to the sign of the Jarlskog invariant $J_{\rm MNS}$~(\ref{JMNS}). 
 
The remaining three parameters $s_{l}, s_{l'}, $ and $s_{\n}$ are determined 
by comparing with observations. 
The PDG parameterization of the mixing matrix is given by 
\begin{align}
U_{\rm MNS} = 
U( \th_{23}^{\rm MNS}, \th_{12}^{\rm MNS}, \th_{13}^{\rm MNS}, \d^{\rm MNS})  ~ 
 {\rm diag} (1 ,  e^{ i \a_{12} / 2} , e^{ i \a_{13} / 2}) ,
\label{PDG}
\end{align}
where $\a_{12}, \a_{13}$ are the Majorana phases, and $U$ is the unitary matrix that is a function of mixing angles $\th_{ij}^{\rm MNS}$ and the Dirac phase $\d^{\rm MNS}$. 
The latest best fit of the normal ordering\footnote{As will be shown later, the inverted ordering is inconsistent with the three-zero texture and DRS.} in the PDG parameterization are given by \cite{Esteban:2020cvm};
\begin{align}
\th_{23}^{\rm MNS} / ^{\circ} &= 49.2^{+0.9}_{-1.2} , ~~~
\th_{12}^{\rm MNS} / ^{\circ} = 33.44^{+0.77}_{-0.74} , ~~~
\th_{13}^{\rm MNS} /^{\circ} = 8.57^{+0.12}_{-0.12} , ~~~
\d^{\rm MNS} / ^{\circ} = 197^{+27}_{-24} . 
\label{35}
\end{align}
From Eq.~(\ref{35}), each parameter is calculated to be 
\begin{align}
s_{l} =  0.7577, ~~
s_{l'}=  \mp \, 0.1402, ~~
s_{\n}= \pm \, 0.5500 . 
\label{sines}
\end{align}
where the upper sign corresponds to positive $s_{e} > 0$. 
These signs are chosen to reproduce the Jarlskog invariants~(\ref{JMNS}).
Unlike the case of CKM matrix~(\ref{Jarls}), the signs of the eight parameters $c_{f}, s_{f}$ in Eq.~(\ref{UMNS1}) cannot be eliminated by redefinition of the five lepton fields. 
Thus, it is necessary to fix the sign of $\cos \d_{\rm MNS}$.
In order to keep $\cos \d_{\rm MNS} < 0$, we impose $s_{l} > 0$. 

A reconstructed MNS matrix~(\ref{UMNS1}) is found to be
\begin{align}
 U_{\rm MNS}' & = O_{e}^{T} \Diag{+i}{1}{1} O_{\n}
\simeq
\begin{pmatrix}
i \, c_{e} & - s_{e} & 0 \\
i \, s_{e} & c_{e} & 0 \\
 0 & 0 & 1 \\
\end{pmatrix} 
\begin{pmatrix}
 0.8269 & \pm 0.5446 & \mp 0.1402 \\
 \mp 0.2702 & 0.6034 &  0.7503 \\
 \pm 0.4932 & - 0.5825 & 0.6461 \\
\end{pmatrix} . 
\label{UMNS3}
\end{align}
Its absolute value is
\begin{align}
|U_{\rm MNS}^{'}|
&= 
\begin{pmatrix}
0.8251 & 0.5449 & 0.1490 \\
0.2755 & 0.6031 & 0.7485 \\
0.4932 & 0.5825 & 0.6461 
\end{pmatrix}  
\label{UMNS2}  . 
\end{align}
In Eq.~(\ref{UMNS2}), dependence of the sign of $s_e$ does not exist.
The errors between the best fit are calculated as
\begin{align}
|U_{\rm MNS}| - |U_{\rm MNS}'| =
\begin{pmatrix}
0.0 & 0.0 & 0.0 \\
4.1 & -1.8 & 0.0 \\
-2.2 & 1.9 & 0.0
\end{pmatrix} \times 10^{-3}  ,  
\end{align}
which are in the range of $O (10^{-3})$.

Next, $CP$ phases will be evaluated.
The Jarlskog invariant gives the Dirac phase $\d^{\rm MNS}$ as
\begin{align}
J_{\rm MNS} &= {\rm Im} \, [ U_{\a i} U_{\b j} U_{\a j}^{*} U_{\b i}^{*} ] \\
&=  \sin \d^{\rm MNS} \,  s_{12}^{\rm MNS} c_{12}^{\rm MNS} s_{13}^{\rm MNS} (c_{13}^{\rm MNS})^{2} s_{23}^{\rm MNS}  c_{23}^{\rm MNS} 
\\ & \simeq -  s_{e} c_{e} c_{l}  c_{l'} c_{\n} s_{\n} s_{l}^2 . 
\label{JMNS}
\end{align}
Calculating the invariant from Eq.~(\ref{UMNS3}) produces
\begin{align}
J_{\rm MNS} = -0.0127, ~~~ 
\sin \d^{\rm MNS} &= -0.381 \simeq + s_{e} \, c_{e} \, s_{l} / s_{l'} \, c_{l'}. 
\label{invJ}
\end{align}
The signs of $J_{\rm MNS}$ and $\sin \d^{\rm MNS}$ do not depend on the sign of $s_{l}$.
Since the sign of $s_{e}$ and $s_{l'}$ are opposite in Eq.~(\ref{sines}), 
the correct sign of the invariant is obtained.

Because $\cos \d^{\rm MNS} < 0$ can be shown similarly, the Dirac phase is found to be $202^{\circ}$ \cite{Yang:2020qsa, Yang:2020goc}. 
Since errors in the mixing angles~(\ref{35}) is at most 4\% in the $1 \s$ regions, the prediction is  expected to have errors of 4$\sim$5 \%.
\begin{align}
\d^{\rm MNS}/^{\circ} \simeq 180 + 22^{+1}_{-1} \,  .
\end{align}
This is very close to the recent best fit for the normal ordering 
$\d^{\rm MNS} / ^{\circ} = 197^{+27}_{-24}$ \cite{Esteban:2020cvm}.

The Majorana phases $\a_{12}$ and $\a_{13}$ can be evaluated by similar rephasing invariants \cite{Branco:1986gr, Jenkins:2007ip, Branco:2011zb}
\begin{align}
\a_{12} = {\rm Arg} \, I_{1} & = {\rm Arg} [ (U_{\rm MNS})_{12}^{2} (U_{\rm MNS})_{11}^{*2} ] \, , 
\label{I1} \\
\a_{13}' = {\rm Arg} \, I_{2} & ={\rm Arg} [ (U_{\rm MNS})_{13}^{2} (U_{\rm MNS})_{11}^{*2}] \, , 
\label{I2}
\end{align}
where $\a'_{13} \equiv \a_{13} - 2 \, \d^{\rm MNS}$. 
Substituting Eq.~(\ref{UMNS3}) into Eqs.~(\ref{I1}) and (\ref{I2}),  we obtain
\begin{align}
& \a_{12}^{0} / ^{\circ} \simeq 11.3^{+0.6}_{-0.6}, ~~~ \a_{13}^{0} / ^{\circ} \simeq 6.90^{+0.4}_{-0.4} .  
\label{alpha0}
\end{align}

Since Eq.~(\ref{UMNS3}) and (\ref{alpha0}) do not count contributions from phases of mass eigenvalues, these effects are parameterized as
\begin{align}
m_{\n 2} = e^{i \b_{2}} m_{2} , ~~~ m_{\n 3} = e^{i \b_{3}} m_{3} . 
\end{align}
Because the symmetries~(\ref{diagref}) fix the phases $\b_{2,3}$ to be $0$ or $\pi$, 
the full Majorana phases are sum of $\a^{0}_{ij}$ and $\b_{j}$;
\begin{align}
(\a_{12}, \, \a_{13}) &= (\a_{12}^{0} + \b_{2}, \, \a_{13}^{0} + \b_{3}) \nn \\
&= (11.3^{\circ}~{\rm or}~ 191.3^{\circ}, ~ 6.90^{\circ}~{\rm or}~ 186.9^{\circ}). 
\end{align}
%

\subsection{Reconstruction of mass matrices}

Here, we reconstruct mass matrices of leptons with three-zero texture and DRS~(\ref{threezero}). 
Although this system has nine parameters, 
there are only eight input parameters in the lepton sector (three charged lepton masses, two neutrino mass differences, and three mixings). 
Due to this, one physical observable (in this case $m_{1}$) remains to be undetermined.
Thus, we will reduce the number of parameters by imposing further $d$\,-\,$e$ unified relation  on the charged lepton sector.

The charged lepton masses at scale $m_{Z}$ are given by \cite{Xing:2011aa}
\begin{align}
m_{e} &= 486.570 \, [\keV],  ~~~
m_{\m} = 102.718 \, [\MeV], ~~~
m_{\t} = 1746.17 \, [\MeV]. 
\end{align}
The mass matrix of charged leptons $M_{e}$ can be made to have the same texture~(\ref{threezero}) as the down type quark.
By redefining fields, we can make $A_{e} > 0$ and $C_{e} > 0$ without loss of generality. 
In this basis, signs of mass eigenvalues relate to mixing angle because $ {\rm sign} ( C_{e} /\tilde B_{e}) =  {\rm sign} (\tilde B_{e}) =  {\rm sign} ( m_{e2} ) = - {\rm sign} ( m_{e1} )$. 
Multiplying the orthogonal matrix $O_{e}$~(\ref{Ue}) generates a Hermitian matrix $M_{e}$ as follows 
\begin{align}
M_{e} = O_{e} \Diag{ \mp m_{e}}{\pm m_{\m}}{m_{\t}} O_{e}^{T} 
 \simeq 
\begin{pmatrix}
0 & \sqrt{m_{e} m_{\m} } & 0 \\
\sqrt{m_{e} m_{\m} } & \pm m_{\m} + s_{\t}^{2} m_{\t} &  s_{\t} m_{\t} \\
0 & s_{\t} m_{\t} & m_{\t}
\end{pmatrix} ,
\end{align}
where these signs are the same as in Eq.~(\ref{sines}). 
The different sign of $s_{e}$ does lead to different textures. 
Interestingly, 
according to the generalized $b$\,-\,$\t$ unification  \cite{Morisi:2011pt, CentellesChulia:2017koy} 
that is valid with an accuracy of about 10\%,
\begin{align}
{ \sqrt{m_{d} m_{s}} / m_{b} } \simeq { \sqrt{m_{e} m_{\m}} / m_{\t} } \, , 
\end{align}
the following relation holds for these textures at the weak scale
\begin{align}
(M_{d})_{12} / (M_{d})_{33}  \simeq (M_{e})_{12} / (M_{e})_{33} \, . 
\end{align}

The parameter $s_{\t}$ is undetermined, as is the 23 mixing of down quark $s_{b}$ in Eq.~(\ref{Ud}). The  $s_b$ is considered to be at most about $|V_{cb}| \sim 0.05$. 
As a comparable value, here $s_{\t}$ is assumed to move in a range $s_{\t}\in [-0.1, \, 0.1]$. 
Once the texture of $M_{e}$ is given, five observables of neutrinos determine five free parameters. 
The mass matrix of neutrinos $M_{\n}$ can be reconstructed by $O_{\n}$ (\ref{Un}) and $s_{l}, s_{l'}, s_{\n}$ (\ref{sines}); 
\begin{align}
M_{\n} = 
U_{\n} \Diag{m_{1}}{m_{2} e^{i \b_{2}}}{m_{3} e^{i \b_{3}}} U_{\n}^{T}  , ~~~ 
U_{\n} = \Diag{+i}{1}{1} O_{\n} . 
\label{fullmn}
\end{align}
Here,  $m_{2,3}$ are functions of $m_{1}$ 
\begin{align}
m_{2} = \sqrt{\D m_{21}^{2} + m_{1}^{2}} \, , ~~~
m_{3} = \sqrt{\D m_{31}^{2} + m_{1}^{2}} \, , 
\end{align}
through the mass-squared differences for the normal ordering \cite{Esteban:2020cvm}
\begin{align}
\D m_{21}^{2} &= 74.2 \, [\meV^{2}], ~~~ 
\D m_{31}^{2} =  2517 \, [\meV^{2}]. 
\end{align}
The 13 matrix element of Eq.~(\ref{fullmn}) is
\begin{align}
(M_{\n})_{13} &= m_{1} U_{\n 11} U_{\n 31} + 
e^{i \b_{2}} m_{2} U_{\n 12} U_{\n 32} +e^{i \b_{3}} m_{3} U_{\n 13} U_{\n 33} \, . 
\label{m13}
\end{align}
Since $U_{\n}$ is a function of $s_{\t}$, 
imposing Eq.~(\ref{m13}) to be zero predicts a value of $m_{1}$ for a given $s_{\t}$.
Therefore, $s_{\t}$ or $m_{1}$ is the only free parameter. The relationship between them is shown in Fig. 1.
This result does not depend on the sign of $s_{e}$.
\begin{figure}[h]
\begin{center}
   \includegraphics[width=8cm]{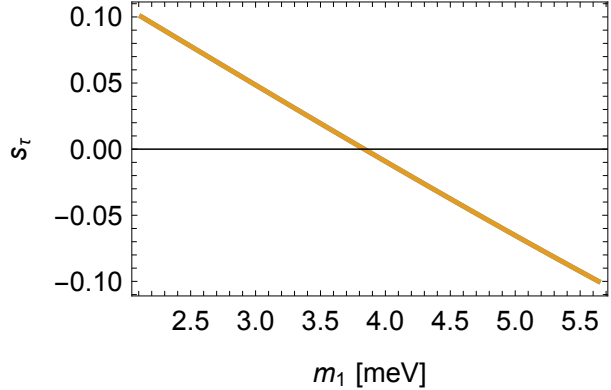} 
\caption{The relation between the lightest neutrino mass $m_{1}$ and the 23 mixing $s_{\t}$ of the matrix $O_{e}$ that diagonalizes the charged lepton mass $M_{e}$.}
\end{center}
\end{figure}
From Figure 1, we can see a range of $m_{1}$ such that $s_{\t}$ is retained a small mixing $|s_{\t}| \lesssim 0.1$.
\begin{align}
& (M_{\n})_{13}  = 0 , ~ s_{\t}\in [-0.1, \, 0.1]  \nn \\
& \To ~ \b_{2} = \pi, ~ \b_{3} = 0, ~  m_{1} = 2.12 \sim 5.64 \, [\meV] \, . 
\end{align}
A solution with $\b_{2} = \b_{3} = 0$ is excluded because it requires large $s_{\t} \sim 1$.
The inverted ordering contradict the zero texture condition $(M_{\n})_{13} = 0$. 

The reconstructed mass matrix is found to be 
\begin{align}
M_{\n} =  
\begin{pmatrix}
+ 0.19 & \mp 8.71 \, i  & 0 \\
\mp 8.71 \, i  & 30.5 & 26.0 \\
0 & 26.0 & 13.1 
\end{pmatrix} \, [\meV] \, ,
\label{mn1}
\end{align}
for $m_{1} = 2.12\, [\meV] \, , \, s_{\t} =  +0.1$ and
\begin{align}
M_{\n} = 
\begin{pmatrix}
-1.79 & \mp 10.0 \, i  & 0 \\
\mp 10.0 \, i  & 19.6 & 27.4 \\
0 & 27.4 & 24.5
\end{pmatrix}\, [\meV] \, ,
\label{mn2}
\end{align}
for $m_{1} = 5.64\, [\meV] \, , \, s_{\t} =  -0.1$. 
The sign of $s_{e}$ changes only the signs of $(M_{\n})_{12, 21}$ because this change can be achieved by a phase redefinition of the first generation. 
Eqs.~(\ref{mn1}) and (\ref{mn2}) still satisfy the DRS~(\ref{diagref}). Furthermore, $(M_{\n})_{11} \simeq 0$ approximately implies the universal four-zero texture  
that is realized by $m_{1} \simeq 2.44\,[\meV]$ and $s_{\t} \simeq 0.08$. 
Because the four-zero texture is type-I seesaw invariant \cite{Nishiura:1999yt,Fritzsch:1999ee}, 
it suggests approximate universal four-zero texture for the neutrino Yukawa matrix $Y_{\n}$ and 
the Majorana mass of the right-handed neutrinos $M_{R}$;
\begin{align}
(Y_{\n})_{11}, (Y_{\n})_{13} \simeq 0, ~~~ 
(M_{R})_{11}, (M_{R})_{13} \simeq 0.
\end{align}

The effective mass ${m_{ee}}$ of the double beta decay is also evaluated as 
a function of $m_{1}$.
The evaluation of $|m_{ee}|$ is shown in Fig. 2. 
\begin{figure}[h]
\begin{center}
   \includegraphics[width=8cm]{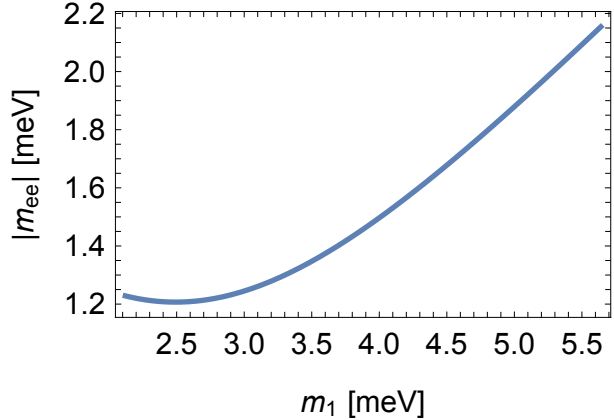} 
\caption{The relation between the lightest neutrino mass $m_{1}$ and the effective mass of double beta decay $|m_{ee}|$.}
\end{center}
\end{figure}
From this, we obtain the following prediction.
\begin{align}
|m_{ee}| &= \left| \sum_{i=1}^{3} m_{i} U_{ei}^{2} \right| \\
&= 1.23 \, [\meV]  ~~ {\rm for} ~~ m_{1} = 2.12 \, [\meV] , \\
 &= 2.15 \, [\meV]  ~~ {\rm for} ~~ m_{1} = 5.64 \, [\meV] . 
\end{align}
Errors of these predicted values come from the input parameters~(\ref{35}) and are considered to be around 3-5\%.

\section{Summary} 

In this paper, we considered a new texture in the SM, 
the three-zero texture of the mass matrices. 
This texture has two less assumptions ($(M_{u})_{11} , (M_{\nu})_{11}\neq 0$) than the  universal four-zero texture $(M_{f})_{11} = (M_{f})_{13,31} =  0$ for $f = u,d,\nu,e$.
The texture allows diagonal reflection symmetries to be almost exact and $d$\,-\,$e$ unification.
They reproduce the CKM and MNS matrices with accuracies of $O(10^{-4})$ and $O(10^{-3})$.
By using a perturbative approximation of the diagonalization, 
two relations with good accuracy for quark masses and mixings are obtained. 

Since this system has nine parameters in the lepton sector, 
in general, the mass of the lightest neutrino $m_{1}$ cannot be predicted.
To reduce a parameter, we assumed that the 23 components of the mass matrices $M_{d,e}$ are comparable.
As new predictions, we obtain the mass of the lightest neutrinos 
 $m_{1} \simeq 2.12\, - \, 5.64\, [\meV]$ and the effective mass of the double beta decay $|m_{ee}| \simeq 1.23 - 2.15 \, [\meV]$. 
Reconstructed neutrino mass exhibits $(M_{\n})_{11} \simeq 0$ and an approximate four-zero texture. 
Because the four-zero texture is type-I seesaw invariant, 
it suggests approximate universal four-zero texture for the neutrino Yukawa matrix $Y_{\n}$ and 
the Majorana mass matrix of the right-handed neutrinos $M_{R}$. 

Even if the basis is changed by a weak basis transformation, other GCPs exist in that basis.  
Since this is just an equivalence transformation, the other GCPs with deformed three-zero texture predict the mixing matrices with high precision in that basis and are almost renormalization invariant.

\section*{Acknowledgment}

This study is financially supported 
by JSPS Grants-in-Aid for Scientific Research
No.~JP18H01210, No. 20K14459,  
and MEXT KAKENHI Grant No.~JP18H05543.


\end{document}